\documentclass{article}
\usepackage{amsmath}
\usepackage{graphicx}
\usepackage{hyperref}
\usepackage{geometry}
\usepackage{authblk}
\usepackage{tikz}

\usetikzlibrary{arrows.meta,positioning,calc,fit}
\usepackage{comment}
\usepackage{tabularx}
\usepackage{amssymb}
\usetikzlibrary{arrows.meta,positioning,calc,decorations.pathmorphing,decorations.markings}
\title{Conceptual Design of a Transverse Deflecting Structure for Longitudinal Diagnostics at DALI }
\author{Najmeh Mirian \thanks{\texttt{najmeh.mirian@hzdr.de}}}
\affil{\small {Helmholtz-Zentrum Dresden-Rossendorf, HZDR, Dresden Germany}}
\date{\today}
\makeatletter         
\def\@maketitle{
\raggedright
\includegraphics[width = 40mm]{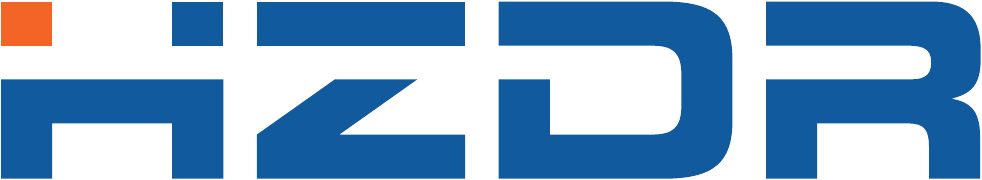}\\[8ex]
\begin{center}
{\Huge \bfseries  \@title }\\[4ex] 
{\Large  \@author}\\[4ex] 
\@date\\[1ex]
\end{center}}
\makeatother
\begin{document}

\maketitle
\tableofcontents
\section{Introduction}

A \textit{Transverse Deflecting Structure} (TDS), also referred to as a \textit{Transverse RF Deflecting Cavity}, is a specialized radio-frequency (RF) traveling-wave cavity designed to generate a transverse electromagnetic force acting on charged particles. The purpose of TDS is 
to impart a time-dependent transverse kick to a charged particle beam. In contrast to accelerating structures, which maximize the longitudinal electric field on axis, the TDS operates in a dipole mode (e.g., TM$_{11}$-type) where the longitudinal electric field varies linearly with transverse displacement and couples the RF fields to the transverse degrees of freedom (typically horizontal or vertical). In other words, TDS enables precise correlation between the particle’s longitudinal position within the bunch and its transverse displacement. As a result, particles experience a transverse Lorentz force that depends on the RF phase. 
When operated near the zero-crossing phase of the deflecting field, the transverse kick varies approximately linearly with the particle arrival time within the bunch. After a drift to an observation screen, this produces a time-to-transverse-position mapping, enabling direct measurement of the longitudinal bunch profile.
\cite{Emma2001, Loew1965RFDeflecting}

In accelerator physics, the primary function of a TDS is to perform \textit{longitudinal phase space diagnostics}. When a bunch traverses a TDS operating at the zero-crossing phase of the RF field, each particle receives a transverse momentum proportional to its arrival time. This temporal-to-spatial mapping allows the downstream observation screen to act as a time-resolved beam monitor. By combining the transverse streak from the TDS with the beam’s energy dispersion in a spectrometer line, one can reconstruct the \textit{full 6D phase space} of the bunch.

%The deflecting performance of the structure is characterized by the transverse shunt impedance, which relates the achievable transverse momentum to the applied RF power and structure parameters. Early implementations and detailed design studies were reported at SLAC in the 1960s \cite{Loew1965RFDeflecting}.
The deflecting performance of an RF transverse deflecting structure is commonly characterized by its transverse shunt impedance, which quantifies the efficiency with which RF power is converted into transverse momentum. For a given input power, structure length, and attenuation, the achievable transverse kick depends directly on this parameter. Optimization of the transverse shunt impedance was a central aspect of early deflector development.

\begin{figure}[t]
\centering
\includegraphics[width=0.7\linewidth]{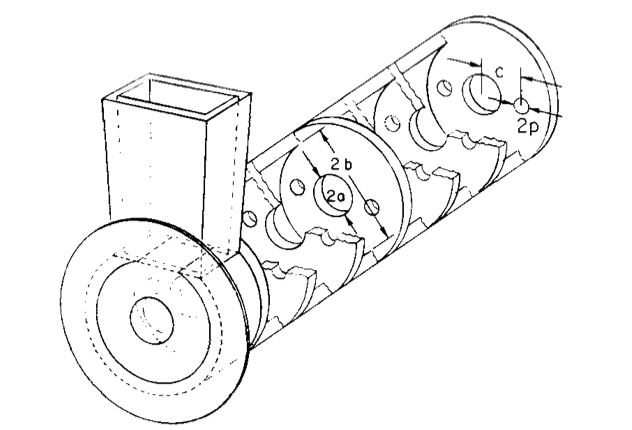}
\caption{TM$_{11}$-type RF deflecting structure (LOLA) developed at SLAC, reproduced from Fig.~1 of \cite{Loew1965RFDeflecting}. The disk-loaded traveling-wave geometry supports a dipole mode used to generate transverse momentum for charged particles.}

\label{fig:tds_lola}
\end{figure}

One of the first systematic implementations was the TM$_{11}$-type disk-loaded traveling-wave structure developed in the 1960’s at SLAC, known as LOLA (Longitudinally Loaded Accelerator). This S-band structure operated at 2856 MHz with a phase advance of $2\pi/3$ per cell and was specifically optimized to maximize transverse deflection while maintaining sufficient beam aperture and acceptance. The geometry of the LOLA structure is shown in Fig.~\ref{fig:tds_lola}, illustrating the disk-loaded waveguide configuration and the transverse-mode stabilization holes described in the original SLAC study \cite{Loew1965RFDeflecting}. These structures, with lengths of 2.44 m and 3.66 m, are designed to handle peak input powers of 25–30 MW and provide up to 32 MV of peak transverse deflecting voltage.
The disk-loaded waveguide is constructed from brazed assemblies of precision-machined copper cells, arranged with a periodic spacing of 3.5 cm. The iris radius is 2.24 cm.

\subsection*{Applications}

The main applications of a transverse deflecting structures in accelerator physics include:

\begin{itemize}

    \item \textbf{Bunch length measurement:}
    Direct measurement of the longitudinal charge distribution by mapping time onto transverse position. 
    This enables sub-picosecond and femtosecond-scale characterization of ultra-short electron bunches.

    \item \textbf{Longitudinal phase space reconstruction:}
    When combined with a dispersive section (e.g., dipole spectrometer), the TDS enables full reconstruction 
    of the longitudinal phase space $(z, \delta)$, including correlated energy chirp, nonlinear distortions, 
    and microbunching structures.

    \item \textbf{Slice emittance and slice energy spread measurement:}
    By performing quadrupole scans or optics variation with the TDS streak active, 
    time-resolved slice parameters such as emittance and energy spread can be extracted. 
    These quantities are essential for understanding brightness limitations and compression effects.

    \item \textbf{Compression and microbunching diagnostics:}
    The TDS allows direct observation of compression-induced distortions, space-charge effects, 
    coherent synchrotron radiation (CSR) effects, and microbunching instabilities.

    \item \textbf{FEL and THz source optimization:}
    Time-resolved beam diagnostics provide feedback for optimizing peak current, slice quality, 
    and longitudinal phase space linearization, which are critical for efficient FEL and THz generation.

    \item \textbf{Beam dynamics and wakefield studies:}
    Time-resolved observation of transverse kicks, wakefield-induced distortions, 
    and collective effects within accelerating structures or vacuum components.
    
\end{itemize}

\subsection*{Examples of operational TDS systems}
\begin{itemize}
    \item \textbf{LOLA-type S-band TDS at DESY} (FLASH ): Enables routine sub-10~fs resolution for bunch profile measurements \cite{LOLA2010}.
    \item \textbf{X-band TDS at SLAC} (FACET-II): Provides femtosecond-level temporal resolution for ultra-short bunches in plasma wakefield acceleration studies \cite{Behrens2014}.
    \item \textbf{C-band TDS at SwissFEL and SACLA}: Used for high-resolution longitudinal diagnostics of FEL driver beams \cite{EGO2015381, PSI_CBANC}.
    \item \textbf{PolariX-TDS at PSI and FLASH}: Allows simultaneous measurement of horizontal and vertical slice parameters \cite{Polarix2020,Polarix2024}.
\end{itemize}

Recent advances in RF design, such as higher-frequency operation (X-band and beyond) and dual-mode operation (PolariX-TDS), have further extended the capabilities of TDS systems, enabling sub-femtosecond temporal resolution in state-of-the-art accelerators.
%%%%%%%%%%%%%%%%%%%%%
\section{Theory of Operation}

The TDS operates by exciting an electromagnetic mode in which the dominant field acts transversely to the beam path. Most commonly, a TM$_{110}$-like dipole mode is used, where the transverse electric field $E_\perp$ and longitudinal magnetic field $B_\parallel$ are phased to provide a net transverse momentum kick to relativistic particles \cite{Emma2001, Loew1965RFDeflecting,Krejcik:556141, FLASH2009}.

\subsection{Transverse Kick Mechanism}
When an ultra-relativistic particle bunch passes through a TDS operating at an angular frequency $\omega$, each particle experiences a transverse Lorentz force given by:
\begin{equation}
    F_\perp(t) = q \left[ E_\perp(z,t) + v B_\perp(z,t) \right],
\end{equation}
where $q$ is the particle charge, $v \approx c$ for relativistic beams, and $t$ is the particle’s arrival time with respect to the RF phase.

For a TDS operated at the zero-crossing phase of the RF field, the small kick angle, $\Delta x'\ll 1 $ is linearly proportional to the particle’s longitudinal position $z$ (or time $t$) within the bunch:
\begin{equation}
    \Delta x'(z) \approx \frac{q V_\perp}{p_z c} \sin\left( \frac{\omega z}{c} \right) \approx q V_\perp \frac{\omega z}{p_z c^2},
\end{equation}
where $\omega$  is the angular RF frequency of the deflecting cavity related to the usual RF frequency f, by $\omega=2\pi f$, and $V_\perp$ is the \textit{transverse voltage} of the cavity, defined as:
\begin{equation}
    V_\perp = \int_{0}^{L} \left[ E_\perp(z) + c B_\perp(z) \right] e^{j \frac{\omega z}{c}} \, dz,
\end{equation}
with $L$ being the cavity length. $V_\perp$ is related to the input power P via the empirical relation

\begin{equation}
    V_\perp\approx 5.824 \sqrt{P[MW]}.
\end{equation}

\subsection{Temporal-to-Spatial Mapping}
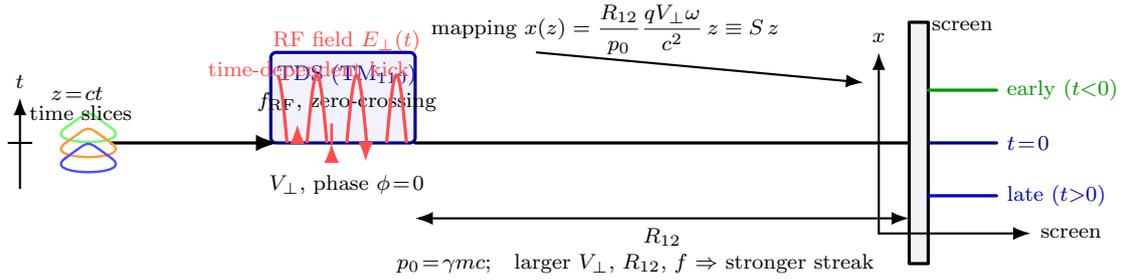
\begin{figure}[h]
\centering
\begin{tikzpicture}[x=1cm,y=1cm,>=Latex,line cap=round,line join=round,thick,
  beam/.style={line width=1.2pt},
  slice/.style={line width=0.9pt},
  lab/.style={font=\footnotesize},
  axis/.style={->,line width=0.8pt},
  scr/.style={fill=gray!10,draw=black},
  cav/.style={fill=blue!5,draw=blue!60!black,very thick,rounded corners=2pt}
]

% --- Layout coordinates ---
\coordinate (src)  at (0,0);
\coordinate (tdsW) at (2.2,0);
\coordinate (tdsE) at (4.0,0);
\coordinate (scr)  at (10.5,0);

% --- TDS cavity block ---
\draw[cav] (tdsW)+(-0.1,1.2) rectangle (tdsE)+(0.1,-1.2);
\node[lab,blue!60!black] at ($(tdsW)!0.5!(tdsE)+(0,0.9)$) {TDS (TM$_{110}$)};
\node[lab] at ($(tdsW)!0.5!(tdsE)+(0,0.55)$) {$f_{\rm RF}$, zero-crossing};
\node[lab] at ($(tdsW)!0.5!(tdsE)+(0,-0.55)$) {$V_\perp$, phase $\phi\!=\!0$};

% --- RF sine inside the cavity (zero-crossing at center) ---
\begin{scope}
\clip (tdsW)+(-0.05,1.15) rectangle (tdsE)+(0.05,-1.15);
\draw[red!70,very thick,domain=2.2:4,samples=120]
  plot(\x,{0.9*sin(12*(\x-3.1) r)});
\end{scope}
\node[lab,red!70] at (3.1,1.35) {RF field $E_\perp(t)$};

% --- Beamline ---
\draw[beam,->] (src) -- (tdsW);
\draw[beam] (tdsE) -- ($(scr)+(0,0)$);

% --- Input bunch (longitudinal slices) ---
\foreach \yy/\clr/\labt in {0.20/green!60/early,0/orange!80/center, -0.20/blue!70/late}{
  \draw[slice,\clr] ($(src)+(-0.6,\yy)$) .. controls + (0.35,0.25) and +(-0.35,0.25) .. ($(src)+(0.0,\yy)$)
                     .. controls + (0.35,-0.25) and +(-0.35,-0.25) .. ($(src)+(-0.6,\yy)$);
}
\node[lab] at (-0.4,0.4) {time slices};
\draw[axis] (-1.2,-0.6) -- (-1.2,0.6) node[above,lab] {$t$};
\draw (-1.35,0) -- (-1.05,0);
\node[lab] at (-0.45,0.65) {$z\!=\!ct$};

% --- Transverse kicks inside the TDS (arrows) ---
\foreach \xx/\amp in {2.45/0.25, 2.9/0, 3.35/-0.25}{
  \draw[->,red!70,line width=1.0pt] (\xx,0) -- ++(0,\amp);
}
\node[lab,red!70] at (2.6,0.95) {time-dependent kick};

% --- Drift label R12 ---
\draw[<->,lab] ($(tdsE)+(0,-1.0)$) -- node[below] {$R_{12}$} ($(scr)+(0,-1.0)$);

% --- Screen ---
\draw[scr,very thick] (scr)+(0, -1.6) rectangle +(0.25, 1.6);
\node[lab] at ($(scr)+(0.7,1.55)$) {screen};

% --- Streak on screen (mapping x = S z) ---
% base line (center slice)
\draw[blue!50!black, very thick]
  ($(scr)+(0.25,0)$) -- ++(0.9,0) node[right,lab] {$t\!=\!0$};

% early slice (positive kick up)
\draw[green!60!black, very thick]
  ($(scr)+(0.25,0.7)$) -- ++(0.9,0) node[right,lab] {early ($t{<}0$)};

% late slice (negative kick down)
\draw[blue!70!black, very thick]
  ($(scr)+(0.25,-0.7)$) -- ++(0.9,0) node[right,lab] {late ($t{>}0$)};

% --- Axes near screen ---
\draw[axis] ($(scr)+(-0.4,-1.2)$) -- ++(0,2.4) node[above,lab] {$x$};
\draw[axis] ($(scr)+(-0.4,-1.2)$) -- ++(2.0,0) node[right,lab] {screen};

% --- Mapping annotation ---
\draw[->,lab] (5.6,1.2) -- (9.9,0.8);
\node[lab] at (6.5,1.5) {mapping $x(z)=\displaystyle \frac{R_{12}}{p_0}\frac{q V_\perp \omega}{c^2}\,z \equiv S\,z$};

% --- Beam energy note ---
\node[lab] at (6.9,-1.6) {$p_0 \!=\! \gamma m c$;\quad larger $V_\perp$, $R_{12}$, $f$ $\Rightarrow$ stronger streak};

\end{tikzpicture}
\caption{Principle of TDS streaking. At RF zero-crossing, the cavity imparts a transverse kick proportional to the arrival time within the bunch. After transport with matrix element $R_{12}$, time is mapped to transverse position on the screen ($x=S\,z$).}
\label{fig:tds_tikz_streak}
\end{figure}

In Fig.~\ref{fig:tds_tikz_streak}, we illustrate the time-to-transverse-position mapping produced by the TDS on the observation screen.
After receiving the transverse kick, the bunch drifts or is transported through downstream optics to an observation screen. The transverse displacement $x$ on the screen for a particle at longitudinal position $z$ is:
\begin{equation}
    x(z) = R_{12} \Delta x',
\end{equation}
where $R_{12}$ is the beam transport matrix element between the TDS and the screen, and $p_0$ is the reference momentum.

This results in a linear mapping:
\begin{equation}
    x(z) = S \cdot z,
    \label{eq:streakstrength}
\end{equation}
where the \textit{streaking strength} $S$ is given by \cite{FLASHTDS2009}:
\begin{equation}
    S = \frac{R_{12}}{p_0} \cdot \frac{q V_\perp \omega}{c^2}.
\end{equation}
%%%%%%%%%%%%%%%%%
%%%%%%%%%%%%%%%%%%%%%%%%%%%%%%%%%%%%%%%%%%%%%%%%%%%%%%%%%%%%%%%%%%%%

%%%%%%%%%%%%%%%%%%%%

%\subsection{Optics Connection of %\texorpdfstring{$R_{12}$}{R12} and Beam Sizes}
%In linear, uncoupled optics, the $2{\times}2$ transport between the TDS location (index $0$) and the screen (index $s$) can be written via Twiss parameters $(\beta,\alpha)$ and phase advance $\Delta\psi$:
%\begin{equation}
%\begin{aligned}
%R_{11} &= \sqrt{\frac{\beta_s}{\beta_0}}\big(\cos\Delta\psi + \alpha_0 \sin\Delta\psi\big),\\
%\boxed{
%\,R_{12} &= \sqrt{\beta_s \beta_0}\,\sin\Delta\psi\, ,\\
%R_{21} &= -\frac{1+\alpha_0\alpha_s}{\sqrt{\beta_s\beta_0}}\sin\Delta\psi
%        + \frac{\alpha_0-\alpha_s}{\sqrt{\beta_s\beta_0}}\cos\Delta\psi,\\
%R_{22} &= \sqrt{\frac{\
\subsection{Optics Connection of $R_{12}$ and Beam Sizes}

In a linear, uncoupled lattice (no coupling, no dispersion terms shown for clarity), the $2\times2$ transverse transport matrix between the TDS location (index 0) and the screen (index s) can be written in terms of Twiss parameters:
\begin{equation}
\begin{pmatrix} x \\ x' \end{pmatrix}_{\!s}
=
\begin{pmatrix}
R_{11} & R_{12} \\
R_{21} & R_{22}
\end{pmatrix}
\begin{pmatrix} x \\ x' \end{pmatrix}_{\!0}, \qquad
\end{equation}
\begin{equation}
\begin{aligned}
R_{11} &= \sqrt{\frac{\beta_s}{\beta_0}}\Big(\cos\Delta\psi + \alpha_0 \sin\Delta\psi\Big),\\
R_{12} &= \sqrt{\beta_s \beta_0}\,\sin\Delta\psi,\\
R_{21} &= -\frac{1+\alpha_0\alpha_s}{\sqrt{\beta_s\beta_0}}\sin\Delta\psi
        + \frac{\alpha_0-\alpha_s}{\sqrt{\beta_s\beta_0}}\cos\Delta\psi,\\
R_{22} &= \sqrt{\frac{\beta_0}{\beta_s}}\Big(\cos\Delta\psi - \alpha_s \sin\Delta\psi\Big),
\end{aligned}
\label{eq:R12_twiss}
\end{equation}
where $\beta_0, \alpha_0$ are the Twiss parameters at the TDS, $\beta_s, \alpha_s$ at the screen, and $\Delta\psi$ is the betatron phase advance between them. Equation~\eqref{eq:R12_twiss} shows explicitly how optics choices set $R_{12}$.

\paragraph{Streaking strength in Twiss form.}
Inserting $R_{12}=\sqrt{\beta_s\beta_0}\sin\Delta\psi$ into the streaking strength $S$ gives
\begin{equation}
S \;=\; \frac{q V_\perp \omega}{c^2\,p_0}\, \sqrt{\beta_s\beta_0}\, \sin\Delta\psi.
\label{eq:S_twiss}
\end{equation}
Hence, for fixed RF/beam parameters, $S$ scales with $\sqrt{\beta_s\beta_0}$ and is maximized near $\Delta\psi \simeq \pi/2$.

\paragraph{Unstreaked size on the screen.}
The rms beam size on the screen without TDS deflection is
\begin{equation}
\sigma_{x,0}^2 \;=\; \varepsilon_x\,\beta_s \;+\; \eta_s^2\,\sigma_\delta^2,
\label{eq:sigx0}
\end{equation}
where $\varepsilon_x$ is the (geometric) emittance, $\eta_s$ the horizontal dispersion at the screen, and $\sigma_\delta$ the relative momentum spread. In a pure time-resolving (non-spectrometer) setup, one tunes $\eta_s \!\approx\! 0$ to minimize $\sigma_{x,0}$, yielding $\sigma_{x,0}\!=\!\sqrt{\varepsilon_x\beta_s}$.

The beam size in the TDS itself is
\begin{equation}
\sigma_{x,{\rm TDS}}^2 \;=\; \varepsilon_x\,\beta_0 \;+\; \eta_0^2\,\sigma_\delta^2,
\end{equation}
which matters for aperture, wakefields, and field linearity in the cavity (keeping $\sigma_{x,{\rm TDS}}$ moderate helps).

\subsection{Temporal Resolution}
Defining temporal resolution via $\sigma_t^{\mathrm{res}} = \sigma_{x,0}/|S|$, and using Eqs.~\eqref{eq:S_twiss}–\eqref{eq:sigx0}, one finds the general Twiss form
\begin{equation}
    \sigma_t^{\mathrm{res}}
    \;=\;
    \frac{c\,p_0}{q V_\perp \omega}
    \;\frac{\sqrt{\varepsilon_x\,\beta_s + \eta_s^2 \sigma_\delta^2}}
    {\sqrt{\beta_s\beta_0}\,|\sin\Delta\psi|}.
    \label{eq:timeres_twiss_general}
\end{equation}
With $\eta_s\!\approx\!0$ (recommended in the streak plane),
\begin{equation}
    \boxed{\;
    \sigma_t^{\mathrm{res}}
    \;=\;
    \frac{c^2\,p_0}{q V_\perp \omega}
    \;\frac{\sqrt{\varepsilon_x/\beta_0}}{|\sin\Delta\psi|}
    \;}
    \label{eq:timeres_twiss_simple}
\end{equation}
which makes the optics levers explicit: smaller $\beta_0$ at the TDS and $\Delta\psi\!\to\!90^\circ$ improve resolution.

\begin{center}
\noindent\fbox{%
\parbox{0.94\linewidth}{%
\textbf{Rule of thumb for best TDS resolution:}
\quad $\Delta\psi \approx 90^\circ$ between TDS and screen,\;
small $\beta_0$ at the TDS in the streak plane,\;
$\eta_s \approx 0$ in the streak plane.\;
Increase $V_\perp$ and $f_{\rm RF}$ to strengthen $S$; higher $p_0$ makes resolution harder.
}}
\end{center}

%\subsection{Extended Capabilities}
Combining the time-dependent streak with an energy-dispersive section (e.g., a dipole spectrometer) enables longitudinal phase-space reconstruction. Advanced designs with dual-mode or dual-plane TDS allow simultaneous streaking in both planes and, together with dispersion, enable slice-resolved 6D diagnostics.

%**************

%%%%%%%%%%%%%%%%%%%%%%%%%%%%%%%%%%%%%%%%%%%%%%%%
\begin{figure}[h]
\centering
\begin{tikzpicture}[x=1cm,y=1cm,>=Latex,line cap=round,line join=round,thick,
  beam/.style={line width=1.2pt},
  slice/.style={line width=0.9pt},
  lab/.style={font=\footnotesize},
  axis/.style={->,line width=0.8pt},
  block/.style={very thick,rounded corners=2pt},
  cav/.style={block,fill=blue!5,draw=blue!60!black},
  dip/.style={block,fill=orange!8,draw=orange!70!black},
  scr/.style={fill=gray!10,draw=black}
]

% --- Key coordinates ---
\coordinate (src)  at (0,0);
\coordinate (tdsW) at (2.0,0);
\coordinate (tdsE) at (3.6,0);
\coordinate (dipC) at (6.2,0.8);  % center of dipole arc
\coordinate (scr)  at (10.5,-0.2);

% --- Incoming beam ---
\draw[beam,->] (src) -- (tdsW);

% --- TDS cavity ---
\draw[cav] (tdsW)+(-0.1,1.2) rectangle (tdsE)+(0.1,-1.2);
\node[lab,blue!70!black] at ($(tdsW)!0.5!(tdsE)+(0,0.9)$) {TDS (TM$_{110}$)};
\node[lab] at ($(tdsW)!0.5!(tdsE)+(0,0.55)$) {$\phi=0$ (zero-crossing)};
\node[lab] at ($(tdsW)!0.5!(tdsE)+(0,-0.55)$) {$V_\perp$, $f_{\rm RF}$};

% RF sine inside TDS (visual)
\begin{scope}
\clip (tdsW)+(-0.05,1.15) rectangle (tdsE)+(0.05,-1.15);
\draw[red!70,very thick,domain=2.0:3.6,samples=120]
  plot(\x,{0.9*sin(12*(\x-2.8) r)});
\end{scope}
\node[lab,red!70] at (1.7,1.35) {time-dependent kick};

% --- Drift to dipole (R12) ---
\draw[beam] (tdsE) -- (4.7,0);
\draw[<->,lab] ($(tdsE)+(0,-1.0)$) -- node[below] {$R_{12}$} (4.7,-1.0);

% --- Sector dipole spectrometer ---
% Entry/exit straight guides
\draw[beam] (4.7,0) -- (5.3,0);
\draw[dip] (5.3,0.9) arc[start angle=90,end angle=15,radius=0.9];
\draw[dip] (5.3,-0.9) arc[start angle=270,end angle=345,radius=0.9];
% Beam trajectory through dipole (bending downward with dispersion)
\draw[beam,postaction={decorate,decoration={markings,mark=at position 0.6 with {\arrow{Latex}}}}]
  (5.3,0) .. controls +(0.9,-0.1) and +(-0.5,0.2) .. (7.2,-0.5) -- (9.2,-0.5);

\node[lab,orange!70!black] at (6.2,1.25) {Spectrometer dipole};
\node[lab] at (6.2,-1.25) {dispersion $D$ ($R_{16}$ or $R_{26}$)};

% --- Screen ---
\draw[scr,very thick] (scr)+(-0.05, -1.6) rectangle +(0.23, 1.6);
\node[lab] at ($(scr)+(0.7,1.55)$) {screen};

% --- Streaking axes near screen ---
\draw[axis] ($(scr)+(-0.8,-1.2)$) -- ++(0,2.4) node[above,lab] {$y$ (energy)};
\draw[axis] ($(scr)+(-0.6,-1.2)$) -- ++(2.0,0) node[right,lab] {$x$ (time)};

% --- Mapping annotations ---
\node[lab,align=left] at (3.0,-1.7)
  {TDS: $x = \underbrace{\frac{R_{12}}{p_0}\frac{q V_\perp \omega}{c}}_{S}\, z \;\; \Rightarrow \;\; x \propto t$};
\node[lab,align=left] at (8.0,-1.7)
  {Dipole: $y = D\,\delta$ \quad with $\delta=\frac{\Delta p}{p_0}$};

% --- Example slices with time colors before TDS ---
\foreach \yy/\clr/\lbl in {0.18/green!60/early,0/orange!80/mid,-0.18/blue!70/late}{
  \draw[slice,\clr] ($(src)+(-0.6,\yy)$) .. controls + (0.35,0.25) and +(-0.35,0.25) .. ($(src)+(0.0,\yy)$)
                     .. controls + (0.35,-0.25) and +(-0.35,-0.25) .. ($(src)+(-0.6,\yy)$);
}

% --- Kicks at TDS (arrows) ---
\foreach \xx/\amp in {2.2/0.25, 2.8/0, 3.3/-0.25}{
  \draw[->,red!70,line width=1.0pt] (\xx,0) -- ++(0,\amp);
}
\node[lab,red!70] at (1.,1.05) {time $\to$ $x$};

% --- Energy-dependent vertical separation at screen ---
\draw[blue!70!black, very thick]  ($(scr)+(0.25, 0.7)$) -- ++(0.9,0) node[right,lab] {$\delta>0$ (higher $p$)};
\draw[blue!50!black, very thick]  ($(scr)+(0.25, 0.0)$) -- ++(0.9,0) node[right,lab] {$\delta=0$};
\draw[blue!40!black, very thick]  ($(scr)+(0.25,-0.7)$) -- ++(0.9,0) node[right,lab] {$\delta<0$ (lower $p$)};

% --- Legend box ---
\draw[rounded corners=2pt,fill=white,draw=black] (0.2,1.5) rectangle (4,2.5);
\draw[beam] (0.2,2.2) -- ++(0.5,0); \node[lab] at (1.6,2.2) {beam path};
\draw[cav] (2.5,2.) rectangle (2.7,2.25); \node[lab] at (3.2,2.25) {TDS};
\draw[dip] (0.55,1.8) arc[start angle=90,end angle=10,radius=0.25]; \node[lab] at (1.45,1.8) {dipole};
%\draw[scr] (2.3,2.2) rectangle (2.41,2); \node[lab] at (2.8,1.8) {screen};

\end{tikzpicture}
\caption{Longitudinal phase-space measurement with a TDS and a spectrometer dipole. The TDS maps arrival time to horizontal position ($x\!\propto\!t$), while the dipole introduces vertical dispersion so that $y\!\propto\!\delta$. The image on the screen encodes $(t,\delta)$ for each slice, enabling reconstruction of the longitudinal phase space.}
\label{fig:tds_spectrometer_phase_space}
\end{figure}
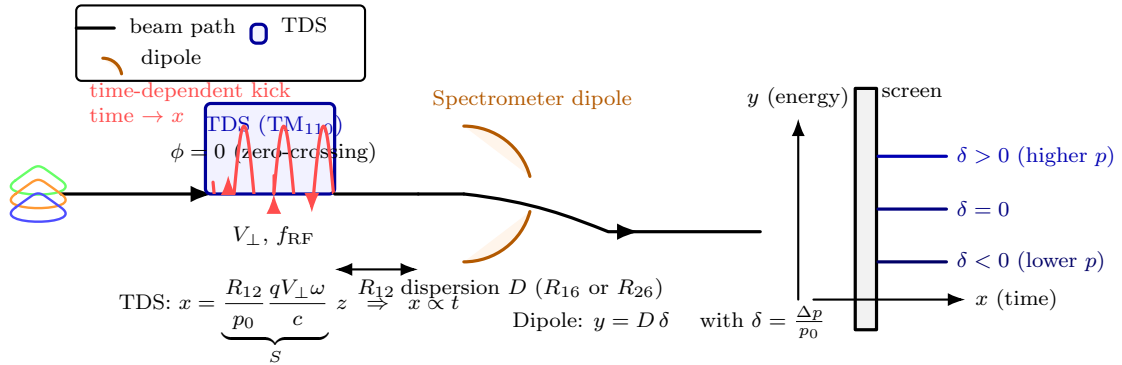
\subsection{Slice Energy and Slice Energy-Spread Measurement: Resolution and Error Sources}

Figure~\ref{fig:tds_spectrometer_phase_space} illustrates the principle of longitudinal phase-space reconstruction using a transverse deflecting structure combined with a spectrometer dipole. 
The TDS converts the longitudinal coordinate into a transverse displacement ($x \propto t$), while vertical dispersion $D \equiv \eta_y$ maps relative momentum deviation $\delta=\Delta p/p$ into vertical position. 
The recorded 2D screen image therefore represents the longitudinal phase space distribution $f(t,\delta)$.

In this configuration,
\begin{itemize}
    \item The horizontal axis encodes arrival time within the bunch:
$x = S\,c\,t ,$
    \item The vertical axis encodes relative energy deviation:
    $
    y = D\,\delta .$ %where $\delta=\Delat p/p$ and $D$ is the vertical dispersion generated by the dipole. 
\end{itemize}
The vertical position on the screen depends linearly on the momentum deviation.
\paragraph{Slice quantities.}
For each time slice $t_i$ (horizontal bin), the mean slice energy and slice energy spread are obtained from the vertical distribution:

\begin{align}
E_{\mathrm{slice}}(t_i) &= E_0\left[1+\langle \delta \rangle_{t_i}\right], 
\qquad
\langle \delta \rangle_{t_i} = \frac{\langle y \rangle_{t_i}}{D},\\
\sigma_{\delta}(t_i) &= \frac{\sigma_y(t_i)}{D}.
\end{align}

Thus, slice energy information is extracted directly from the centroid and width of the vertical projection for each time bin, $\sigma_E=\sigma_y(t_i) E/D$.

\paragraph{Intrinsic resolution limit.}
%The smallest measurable slice energy spread is fundamentally limited by the vertical beam size in the absence of dispersion. Even for a monoenergetic beam, the vertical image contains contributions from the optical beta-function and the imaging system resolution:
In a dispersive region of the beamline, the vertical beam size can be expressed as the quadratic sum of two components: the dispersion-induced contribution associated with the relative energy spread $D\sigma_E/E$, and the intrinsic betatron beam size $\sqrt{\frac{m_e c^2}{E}\,\beta_y\,\varepsilon_{ny}}$. In addition to the dispersive and intrinsic beam size components, the observed beam size will be affected by the profile monitor resolution $\sigma_r$, which is assumed to be independent on the energy. Assuming that the dispersion-induced term, the intrinsic beam size, and the monitor resolution are uncorrelated, the measured beam size% $\sigma_y$:
	
  is given by the quadratic sum of these contributions.
\begin{equation}
\sigma_{y}^2 = \sigma_R^2 + \frac{m_e c^2}{E}\,\beta_y\,\varepsilon_{ny} +\frac{D^2\sigma_E^2}{E^2}.
\end{equation}

This expression defines the resolution floor: increasing dispersion $D$ improves energy resolution, whereas large $\beta_{y,s}$ or imaging blur $\sigma_R$ degrade it.

\paragraph{TDS-induced energy spread.}
In addition to measurement limitations, the deflecting cavity itself can introduce uncorrelated energy spread due to transverse-longitudinal coupling \cite{Floettmann}:

\begin{equation}
\sigma_E^{\mathrm{TDS}} = q\,k\,V_\perp\,\cos\phi\;\sigma_T,
\qquad
\sigma_T = \sqrt{\frac{m_e c^2}{E}\,\beta_T\,\varepsilon_{nT}}.
\end{equation}
where k is the RF wave number of the deflector,  $\phi$ is the operating phase of the deflector (normally the deflector operates at the zero), and $\sigma_T$ is the average transverse beam size in the structure, which can be obtained from the beat-function, the emittance in the streaking plane and the energy at the deflector (respectively $\beta_T$, $\varepsilon_{nT}$ , and $E$). For slice energy spread measurements, the deflecting voltage must be carefully optimized to balance two competing effects: minimizing the additional energy spread induced by the cavity while providing sufficient streaking strength to resolve the longitudinal structure and mitigate RF curvature effects.

%where $\beta_T$ is the $\beta$-function at the profile monitor, $\varepsilon_{n}$ the normalized emittance.
The energy spread introduced by the deflector is assumed to be uncorrelated with the beam energy spread; therefore

\begin{equation}
\sigma_{y}^2 = \frac{D^2\sigma_E^2}{E^2}+\sigma_R^2 + \frac{m_e c^2}{E}\,\beta_y\,\varepsilon_{ny} +(q\,k\,V_\perp\, )^2\,\cos^2(\phi)\,\frac{m_e c^2}{E}\,\beta_T \,\varepsilon_{nT},
\end{equation}
which must be experimentally separated to recover the true beam properties.

\paragraph{Experimental separation of contributions.}
A practical method to disentangle these effects follows the approach described in~\cite{Eduard2020}. 
By performing an energy scan and measuring the projected vertical beam size $\sigma_M$ for different beam energies.%, one obtains:\\
Repeating the measurement for different TDS voltages allows isolation of $\sigma_E^{\mathrm{TDS}}$ and recovery of the intrinsic slice energy spread $\sigma_{E0}$.

\paragraph{Error sources and systematic limitations.}
The accuracy of slice energy reconstruction depends on multiple coupled effects:

\begin{enumerate}
    \item \textbf{Instrumentation:} screen resolution, camera calibration, and imaging nonlinearity.
    \item \textbf{Optics:} uncertainty in dispersion $D$, $\beta_{y,s}$, and emittance subtraction.
    \item \textbf{TDS stability:} voltage calibration, phase jitter, and wakefield-induced energy modulation.
    \item \textbf{Data analysis:} finite temporal resolution and binning effects.
    \item \textbf{Beam jitter:} arrival-time and energy fluctuations that smear the phase-space image.
\end{enumerate}

These contributions may act simultaneously and must be quantified to achieve reliable slice diagnostics.

\paragraph{Optimization strategy.}
High-resolution slice measurements require maximizing dispersion in the energy plane while minimizing $\beta_{y,s}$ and residual dispersion in the streak plane. 
Systematic calibration of imaging resolution and voltage-dependent scans are essential to isolate instrumental and cavity-induced effects.

\paragraph{Summary box.}
\begin{center}
\noindent\fbox{%
\parbox{0.94\linewidth}{%
\textbf{For best slice energy resolution:}
Low $\beta_{y,s}$, high $D$, accurate $\sigma_R$ calibration, $\eta_x\!\approx\!0$, two-scan method to remove TDS-induced spread, and synchronization to suppress jitter.
}}
\end{center}

%%%%%%%%%%
% ---------- Key relations (boxed) ----------
% ---------- Updated Key relations (TDS time + energy measurement) ----------
\begin{center}
\noindent\fbox{%
\parbox{0.95\linewidth}{%
\textbf{Key relations for TDS time and energy diagnostics:}

\medskip
\underline{\textit{Time-domain (TDS streak plane):}}
\[
S = \frac{R_{12}}{p_0}\,\frac{q V_\perp \omega}{c^2}, 
\quad
x = S\,z = S\,c\,t,
\]
\[
\sigma_t^{\mathrm{res}} = \frac{\sigma_{x,0}}{|S|},
\quad
\sigma_x^2 = \sigma_{x,0}^2 + (S\,c\,\sigma_t)^2.
\]

\underline{\textit{Energy-domain (orthogonal plane with dispersion $D\!=\!\eta_y$):}}
\[
y = D\,\delta,
\quad
\langle \delta \rangle_{t_i} = \frac{\langle y \rangle_{t_i}}{D},
\quad
\sigma_{\delta}(t_i) = \frac{\sigma_y(t_i)}{D}.
\]
\[
\sigma_{\delta}^{\mathrm{res}} = \frac{\sigma_{y,0}}{D},
\quad
\sigma_{y,0}^2 = \sigma_R^2 + \varepsilon_y \beta_{y,s}.
\]

\underline{\textit{TDS-induced uncorrelated energy spread:}}
\[
\sigma_E^{\mathrm{TDS}} = e\,k\,V\,\cos\phi \;\sigma_T,
\quad
\sigma_T = \sqrt{\frac{m_e c^2}{E_T}\,\beta_{T}\,\varepsilon_{n,T}}.
\]
Measured spread adds in quadrature:
\[
\sigma_E^2 = \sigma_{E0}^2 + \bigl(\sigma_E^{\mathrm{TDS}}\bigr)^2.
\]

\underline{\textit{Energy-scan method:}}
\[
\sigma_{y}^2 = \frac{D^2\sigma_E^2}{E^2}+\sigma_R^2 + \frac{m_e c^2}{E}\,\beta_y\,\varepsilon_{ny} +(q\,k\,V_\perp\, )^2\,\cos^2(\phi)\,\frac{m_e c^2}{E}\,\beta_T \,\varepsilon_{nT},
\]
Fit $\sigma_M^2(E)$ at several $E$ to extract $\sigma_R$, $\varepsilon_{n,y}$, and $\sigma_E$.

\medskip
\textit{Note:} $R_{12}=\sqrt{\beta_s\beta_0}\,\sin\Delta\psi$ in Twiss form. Maximize $D$ in energy plane and minimize $\beta_{y,s}$ for best energy resolution.
}%
}
\end{center}
For clarity and quick reference, the key parameters governing TDS diagnostics and their typical numerical scales are summarized in Table~\ref{tab:tds_quickref}. 
These values provide practical context for the resolution limits and optimization strategies discussed above.
As seen in Table~\ref{tab:tds_quickref}, the achievable temporal resolution improves with increasing RF frequency and transverse voltage, while larger transport matrix element $R_{12}$ enhances streaking strength. 
Conversely, dispersion $D$ primarily determines the energy resolution in phase-space measurements. 
The listed numerical ranges reflect typical operating conditions in modern S-band and X-band TDS systems.

%-

% ---------- Quick-reference table ----------
\begin{table}
\centering
\caption{Quick-reference: symbols, meanings, and typical numerical scales in TDS diagnostics.}
\label{tab:tds_quickref}
\begin{tabular}{l l c}
\hline
\textbf{Symbol} & \textbf{Meaning} & \textbf{Typical scale}\\
\hline
$f_{\rm RF}$ & RF frequency of TDS & S-band $2.9$\,GHz;\; X-band $11.4$\,GHz \\
$V_\perp$ & Transverse voltage & S-band $5$–$15$\,MV;\; X-band $30$–$60$\,MV \\
$R_{12}$ & Transport element TDS$\rightarrow$screen & $2$–$10$\,m \\
$D$ & Dispersion at screen (spectrometer) & $0.1$–$0.5$\,m \\
$E$ & Beam energy & injector: $0.1$–$0.3$\,GeV;\; linac: $1$–$17$\,GeV \\
$\sigma_{x,0}$ & Unstreaked rms size on screen & $20$–$200\,\mu$m \\
$\sigma_t^{\mathrm{res}}$ & Temporal resolution (rms) & few fs (S-band) $\to$ sub-fs (X-band) \\
$S$ & Streaking strength & increases with $f_{\rm RF}$, $V_\perp$, $R_{12}$ \\
\hline
\end{tabular}
\end{table}
%%%%%%%%%%%%%%%%%%%5%%%

\section{Design Considerations}

The performance of a TDS is governed by a complex interplay between RF design parameters, beamline optics, and operational stability.  
Temporal resolution, typically quantified through the shear parameter and the achievable streaking strength, ultimately depends on this coupled optimization.
Originally developed as a diagnostic for bunch length measurement in linear accelerators, the TDS has evolved into a high-precision tool capable of mapping the full longitudinal phase space of ultrashort electron bunches. 

Designing a TDS beamline is inherently a balancing act: RF parameters set the streaking strength and resolution limits; beam optics govern the mapping from the cavity to the screen; and operational stability defines whether the theoretical performance can be realized in practice.  
At the same time, higher frequency and stronger streaking improve temporal resolution but demand tighter tolerances, smaller apertures, and greater immunity to wakefields.  
The following discussion examines these trade-offs, outlining how scientific requirements and engineering constraints shape the final design.

\subsection{Cavity Engineering Considerations}

The physical design of a TDS depends strongly on the RF frequency. 
The operating mode is typically the dipole-like \textit{TM$_{110}$} mode, which provides the time-dependent transverse kick. 
Engineering aspects such as cavity diameter, beam aperture, power coupling, and cooling must be carefully optimized to balance 
performance with operational reliability. 

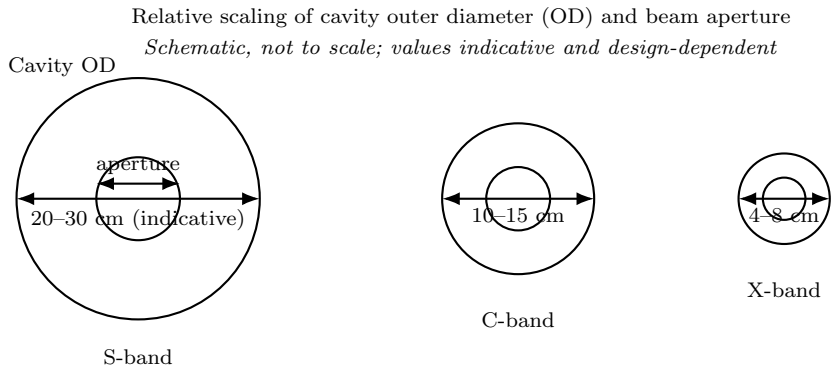
\begin{figure}[h]
\centering
\begin{tikzpicture}[x=1cm,y=1cm,thick,>=Latex,
  lab/.style={font=\footnotesize}]
% S-band (largest)
\begin{scope}[shift={(0,0)}]
  \draw (0,0) circle (1.6);              % outer diameter ~ 32 mm (scaled)
  \draw (0,0) circle (0.55);             % beam aperture
  \node[lab] at (0,-2.1) {S-band};
  \node[lab] at (-1,1.75) {Cavity OD};
  \draw[<->] (-1.6,0) -- (1.6,0) node[midway,below,lab]{20--30 cm (indicative)};
  \draw[<->] (-0.55,0.2) -- (0.55,0.2) node[midway,above,lab]{aperture};
\end{scope}

% C-band (medium)
\begin{scope}[shift={(5,0)}]
  \draw (0,0) circle (1.0);              % outer diameter ~ 20 mm (scaled)
  \draw (0,0) circle (0.42);             % beam aperture
  \node[lab] at (0,-1.6) {C-band};
  \draw[<->] (-1.0,0) -- (1.0,0) node[midway,below,lab]{10--15 cm};
\end{scope}

% X-band (smallest)
\begin{scope}[shift={(8.5,0)}]
  \draw (0,0) circle (0.6);              % outer diameter ~ 12 mm (scaled)
  \draw (0,0) circle (0.28);             % beam aperture
  \node[lab] at (0,-1.2) {X-band};
  \draw[<->] (-0.6,0) -- (0.6,0) node[midway,below,lab]{4--8 cm};
\end{scope}

% Note
\node[lab,align=center] at (4.25,2.2)
  {Relative scaling of cavity outer diameter (OD) and beam aperture \\[2pt]
   \emph{Schematic, not to scale; values indicative and design-dependent}};
\end{tikzpicture}
\caption{Indicative comparison of cavity outer diameter and beam aperture for S-, C-, and X-band TDS cavities. Higher frequency (X-band) yields smaller structures and apertures, which improves streaking efficiency but tightens alignment and increases wakefield sensitivity.}
\label{fig:tds_cavity_cross_sections}
\end{figure}

%\paragraph{Cavity Size and Frequency Dependence.}
The cavity dimensions scale with the RF wavelength $\lambda = c/f$. 
At lower frequencies (S-band), the wavelength is longer, leading to larger cavity diameters and wider apertures, 
which relaxes alignment tolerances but reduces achievable temporal resolution. Figure \ref{fig:tds_cavity_cross_sections} compares cavity outer diameter and beam aperture for S-, C-, and X-band TDS cavities.
The beam aperture must be large enough to pass the full bunch including its halo, while small enough to maintain high transverse 
field strength.
At higher frequencies (X-band), the structures become more compact, support higher gradients, and enable femtosecond-scale resolution, 
but smaller structures and aperture tightens alignment and increases wakefield sensitivity. 
 Smaller apertures in C- and X-band increase the risk of wakefield effects, which are particularly relevant for 
low-energy beams such as DALI (50~MeV). 
At beam energies around 50~MeV, transverse wakefields can induce centroid shifts and slice emittance growth, potentially compromising the very diagnostics the TDS is intended to perform.
%\paragraph{RF Power Coupling and Breakdown.
%%%%%%%%%%%%%%%%%%%%%%%%%%%%%
%%%%%%%%%%%%%%%%%%%%%%%%%%%%%%%%%%%%%%%%%
On the other hand, the cavity requires RF power delivery via input couplers. 
S-band cavities use larger waveguides and can operate robustly at gradients of $\sim 20$--25~MV/m. 
X-band cavities, in contrast, can reach 40--60~MV/m but face higher breakdown probability, requiring careful conditioning and 
high-quality surface finishing. Breakdown probability increases roughly exponentially with surface electric field, making pulse length and surface preparation critical parameters in X-band operation.

%\paragraph{Comparison of S-, C-, and X-band Structures.}
A summary of typical cavity parameters and their associated advantages and disadvantages is given in Table~\ref{tab:tds-bands}.

\paragraph{Cooling and Materials.}
All TDS structures are typically made from Oxygen-free high-conductivity (OFHC) copper with brazed or drilled cooling channels \cite{Wangler2008}. 
Thermal stability is essential, since frequency detuning of even a few tens of kHz can degrade performance. 
Higher frequencies (X-band) dissipate more power in smaller volumes, requiring more efficient cooling solutions. 
The resonant frequency shift is approximately
\[
\Delta f \approx -\alpha f \Delta T,
\]
where $\alpha$ is the thermal expansion coefficient of copper.

\begin{table}[h!]
\centering
\caption{Comparison of cavity characteristics at different frequency bands. 
Values are indicative and depend on detailed design. 
$F$ is the RF frequency, $\lambda$ is RF wavelength, $d$ is cavity diameter, and $G$ is cavity gradient.}
\label{tab:tds-bands}
\begin{tabularx}{\linewidth}{lccccXX}
\hline
Band & $F$ (GHz) & $\lambda$ (cm) & $d$ (cm) & $G$ (MV/m) & Pros & Cons \\
\hline
S-band & $\sim 3$  & 10.0 & 20--30 & 20--25 & Robust, easier cooling & Large size, lower resolution \\
C-band & $\sim 6$  & 5.0  & 10--15 & 30--35 & Balance of size and resolution & More sensitive alignment \\
X-band & $\sim 12$ & 2.5  & 4--8   & 40--60 & High resolution, compact & Tight tolerances, breakdown risk \\
\hline
\end{tabularx}
\end{table}

\subsection*{Operating Frequency, Transverse Voltage and Power}
Modern facilities employ TDS systems across a broad frequency range (S-band to X-band), leveraging advances in RF cavity design, high-power klystron and solid-state amplification, and ultra-stable low-level RF (LLRF) control.
The choice of RF frequency affects both temporal resolution and cavity size:
\begin{itemize}
    \item \textbf{S-band (2--3~GHz):} Suitable for tens-of-femtosecond resolution; larger aperture makes it tolerant to orbit jitter and wakefield effects; lower shunt impedance means higher RF power is required for the same streaking voltage.
    \item \textbf{C-band (5--6~GHz):} Provides higher streaking strength per unit length and better resolution; moderate aperture; still manageable mechanical tolerances.
    \item \textbf{X-band (11--12~GHz):} Enables sub-femtosecond resolution due to high streaking gradient, but requires tight alignment ($<100~\mu$m) and timing stability (few-fs phase jitter). Aperture is smaller, which increases wakefield sensitivity.
\end{itemize}
In practice, frequency selection balances resolution goals against complexity, cost, and tolerance requirements.
 
From the accelerator beamline perspective, the cavity is powered by an external RF source, typically a klystron or solid-state amplifier, and fed through a waveguide coupler designed to excite the deflecting mode with minimal reflection and mode conversion.  
The required peak power depends on the operating frequency band, the cavity length, and its transverse shunt impedance.  

In S-band systems, klystrons in the 10–20 MW class can drive one- to two-meter structures to transverse voltages of several tens of MV, sufficient for tens-of-femtosecond resolution.  
C-band systems, being more compact, achieve higher shunt impedance and therefore require less RF power for the same voltage; typical sources deliver a few MW, often with pulse compression to reach the desired peak.  
X-band cavities push this further: they are physically short with very high gradient potential, enabling sub-femtosecond resolution, but their small apertures and high surface fields demand precise machining, sub-100 µm alignment, and RF sources in the 20–50 MW class.  

From an engineering standpoint, the “entry” to the TDS is therefore twofold: for the beam, it is the entrance aperture and optics that prepare the bunch for streaking; for the RF, it is the power supply and coupler chain that set the achievable transverse kick.  
Both determine whether the theoretical resolution limits can be realized in practice, and both impose practical constraints on cavity design, mechanical stability, and overall beamline integration.

Higher transverse voltage $V_\perp$ increases the streaking strength $S$ and improves time resolution:
\begin{itemize}
    \item For a given frequency, $V_\perp$ is determined by the shunt impedance, cavity length, and RF power.
    \item Standing-wave cavities generally offer higher peak voltage per length, while traveling-wave designs can be shorter with wider bandwidth.
    \item In high-brightness linacs, $V_\perp$ values of 20--40~MV are typical to achieve few-femtosecond resolution.
\end{itemize}
The RF system must deliver stable amplitude and phase to prevent drift in the time-to-position calibration.

\subsection{Wakefields and Beam Loading}

When an electron bunch traverses a transverse deflecting cavity, it excites electromagnetic fields in addition to the externally applied RF deflecting mode. 
These self-induced fields, commonly referred to as wakefields \cite{Wangler2008,ChaoMess2013}, can influence both the transverse and longitudinal beam dynamics and must therefore be carefully considered in the design of a TDS beamline.

\paragraph{Short-Range Wakefields.}
Short-range wakefields arise from the interaction of the bunch with the cavity geometry and are particularly significant in structures with small beam apertures and high operating frequencies. 
In dipole-mode cavities such as the TM$_{110}$ deflecting structure, transverse wakefields can introduce slice-dependent kicks along the bunch length. 
These additional kicks distort the intended linear time-to-transverse mapping \cite{Akre2002} and may lead to artificial growth of slice emittance or centroid shifts at the observation screen.

The strength of short-range wakefields increases as the iris radius decreases, scaling approximately with $1/a^3$ \cite{Palumbo1994}, where $a$ is the aperture radius. 
Consequently, high-frequency cavities (e.g., X-band) with compact geometries are intrinsically more sensitive to wakefield effects. 
For low-energy beams, such as the 50~MeV beam foreseen for DALI, the impact is amplified due to the lower beam rigidity \cite{Lee2018}, making careful aperture optimization essential.

\paragraph{Beam Loading.}
In addition to wakefields, the beam extracts energy from the RF field during its passage through the cavity. 
This phenomenon, known as beam loading \cite{Wangler2008}, reduces the effective transverse deflecting voltage $V_\perp$. 
The reduction scales with bunch charge and cavity shunt impedance and can become significant in high-charge operation.

Beam loading modifies not only the amplitude but also the phase of the effective deflecting field, potentially degrading temporal resolution and introducing systematic measurement errors. 
To mitigate these effects, feedforward systems are often implemented to pre-compensate the RF drive amplitude, while feedback systems can stabilize the cavity field against pulse-to-pulse fluctuations.

\paragraph{Higher-Order Modes (HOMs).}
Beyond the fundamental deflecting mode, the beam excites higher-order modes (HOMs) in the cavity. 
If not sufficiently damped, long-range transverse HOMs can interact with subsequent bunches in multi-bunch operation, leading to cumulative beam breakup (BBU) instabilities or additional trajectory jitter \cite{Lee2018,Palumbo1994}.

Effective HOM damping is therefore an integral part of cavity engineering. 
This typically involves the use of dedicated couplers or waveguide absorbers to extract unwanted modes while preserving the quality factor of the operating deflecting mode. 
In single-bunch operation, HOM effects are less critical but still relevant for preserving beam quality in high-repetition-rate facilities.

\paragraph{Design Implications.}
Short-range wakefields, beam loading, and higher-order mode (HOM) excitation together impose practical constraints on the cavity aperture, operating frequency, and achievable gradient.
Therefore, the TDS design must carefully balance the desired temporal resolution with the sensitivity to wakefield effects. This is especially important in low-energy beamlines, where wakefield-induced transverse kicks can become comparable to the intended RF streaking signal and potentially distort the measurement.

\subsection{Timing Synchronization}

The temporal resolution of a TDS is fundamentally limited not only by streaking strength and beam optics, but also by the synchronization between the RF deflecting field and the electron beam arrival time. 
Since the TDS maps longitudinal time into transverse position through the zero-crossing phase of the RF field, any relative timing jitter directly translates into measurement uncertainty.

\paragraph{RF Phase Stability Requirements.}
The transverse kick imparted by the cavity depends on the instantaneous RF phase at the moment the bunch traverses the structure. 
Operating the TDS near the zero-crossing of the RF waveform ensures a linear time-to-transverse mapping. 
However, phase fluctuations $\Delta \phi$ around this operating point introduce an effective timing error \cite{Wangler2008, CERNRFSchool}

\[
\Delta t = \frac{\Delta \phi}{\omega_{\rm RF}},
\]
where $\omega_{\rm RF} = 2\pi f_{\rm RF}$.

For high-frequency systems, even small phase deviations correspond to femtosecond-scale timing shifts. 
For example, at X-band ($f_{\rm RF} \approx 11.4$~GHz), a phase jitter of 0.1$^\circ$ corresponds to approximately 24~fs of timing error. 
Therefore, achieving sub-femtosecond temporal resolution requires RF phase stability at the level of a few millidegrees, corresponding to timing jitter well below 10~fs.

\paragraph{Sources of Timing Jitter.}
Timing fluctuations arise from multiple sources, including RF oscillator phase noise, thermal drift in RF distribution lines, mechanical length variations, and beam arrival time jitter originating from upstream accelerating structures. 
In low-energy sections, additional contributions may arise from RF amplitude-to-phase conversion and space-charge-induced energy fluctuations, which convert into arrival-time jitter through longitudinal dispersion.

%\paragraph{Optical Timing Distribution Systems.}
%To achieve the synchronization precision required for femtosecond diagnostics, many modern accelerator facilities employ optical timing distribution systems \cite{Schlarb2013}. These systems distribute a stabilized optical reference signal via fiber links to synchronize RF stations and laser systems with sub-10~fs precision.

%Active stabilization techniques compensate for fiber length variations caused by temperature changes and mechanical stress. Balanced optical cross-correlators or phase detectors are commonly used to measure and correct timing drifts in real time. 
%Such systems are routinely implemented at X-ray free-electron laser facilities, where electron beam and laser synchronization at the few-femtosecond level is required.

\paragraph{Design Implications.}
The achievable timing stability directly defines the lower bound of measurable bunch length. 
If the synchronization jitter approaches the intrinsic bunch duration, the measurement becomes dominated by timing noise rather than physical beam structure. 
Consequently, the RF synchronization architecture must be designed in parallel with the TDS cavity and beam optics to ensure that the targeted temporal resolution can be realized under operational conditions \cite{Schlarb2013}.

\subsection{Mechanical Stability and Alignment}
Mechanical stability and alignment accuracy play an equally critical role, particularly in high-frequency systems where tolerances become increasingly stringent.

\paragraph{Transverse Alignment.}
The deflecting field in a TM$_{110}$ cavity provides a transverse kick in a well-defined polarization plane. 
If the cavity is transversely misaligned with respect to the beam trajectory, the beam experiences an unintended transverse offset within the cavity. 
This offset can excite additional dipole fields and couple the streaking signal into the orthogonal plane, thereby distorting the intended time-to-transverse mapping.

Even small misalignments can introduce systematic offsets at the observation screen, which may be incorrectly interpreted as temporal structure. 
For high-resolution diagnostics, the cavity center must therefore be aligned to the beam axis with precision typically better than a fraction of the beam size inside the cavity. 
In practice, beam-based alignment procedures are often required to achieve the necessary accuracy.

\paragraph{Angular Alignment and Roll Errors.}
In addition to transverse displacement, angular misalignment and roll errors of the cavity around the beam axis can degrade measurement quality. 
A roll misalignment rotates the streaking plane, leading to coupling between horizontal and vertical optics and complicating the interpretation of the measured beam profile. 
For femtosecond-scale diagnostics, roll tolerances on the order of a few milliradians or better are generally required.

\paragraph{Vibrations and Mechanical Support.}
Mechanical vibrations of the cavity or its support structure translate directly into transverse position jitter at the screen. 
Since the streaking measurement converts time into transverse position, any mechanical motion becomes indistinguishable from timing jitter. 
This effect is particularly critical in facilities aiming at sub-10~fs resolution.

The cavity support structure must therefore minimize ground vibration coupling and mechanical resonances. 
Rigid supports with high natural frequencies, combined with vibration-damping elements, are typically employed. 
In some cases, active stabilization systems may be considered for ultra-high-resolution applications.

\paragraph{Thermal Stability.}
Thermal expansion of the cavity structure leads to changes in its resonant frequency. 
For copper cavities, the frequency shift scales approximately with the linear thermal expansion coefficient \cite{Behrens2014,Wangler2008},
\[
\frac{\Delta f}{f} \approx -\alpha \Delta T,
\]
where $\alpha \approx 17 \times 10^{-6}~\mathrm{K}^{-1}$ for OFHC copper.

Even temperature variations of a few tenths of a degree can result in frequency shifts of several tens of kilohertz at S-band and proportionally larger shifts at X-band. 
Such detuning modifies the cavity phase relative to the beam arrival time, thereby degrading the streaking calibration and temporal resolution.

To ensure stable operation, temperature control systems typically maintain cavity cooling water stability at the level of $\pm 0.1^\circ$C or better. 
For X-band systems, where frequency sensitivity is higher and apertures are smaller, transverse alignment tolerances below 100~$\mu$m and temperature stability better than $0.1^\circ$C are commonly required.
%%%%%%%%%%%%%%%%%%%%%%%%%

%%%%%%%%%%%%%%%%%%%%%%%%%%%%%%%%%%%%%%%%
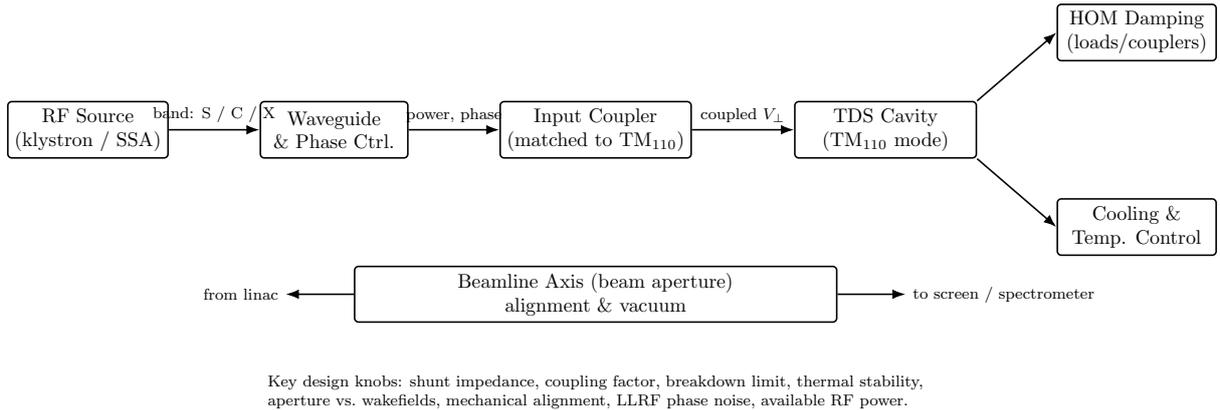
\begin{figure}%[h]
\centering
\resizebox{\linewidth}{!}{%
\begin{tikzpicture}[x=1cm,y=1cm,thick,>=Latex,
  box/.style={draw,rounded corners=2pt,minimum width=2.6cm,minimum height=1.0cm,align=center},
  lab/.style={font=\footnotesize}]

% (your figure code here...)

%\begin{figure}[h]
%\centering
%\begin{tikzpicture}[x=0.6cm,y=0.6cm,thick,>=Latex,
  box/.style={draw,rounded corners=2pt,minimum width=2.4cm,minimum height=1.0cm,align=center},
  lab/.style={font=\footnotesize}]

\node[box] (src) {RF Source\\(klystron / SSA)};
\node[box,right=1.6cm of src] (wg) {Waveguide\\\& Phase Ctrl.};
\node[box,right=1.6cm of wg,minimum width=3.2cm] (cpl) {Input Coupler\\(matched to TM$_{110}$)};
\node[box,right=1.8cm of cpl,minimum width=3.2cm] (cav) {TDS Cavity\\(TM$_{110}$ mode)};
\node[box,above right=0.7cm and 1.4cm of cav,minimum width=2.8cm] (hom) {HOM Damping\\(loads/couplers)};
\node[box,below right=0.7cm and 1.4cm of cav,minimum width=2.8cm] (cool) {Cooling \&\\Temp. Control};

\draw[->] (src) -- node[above,lab]{band: S / C / X} (wg);
\draw[->] (wg) -- node[above,lab]{power, phase} (cpl);
\draw[->] (cpl) -- node[above,lab]{coupled $V_\perp$} (cav);
\draw[->] (cav.north east) -- (hom.west);
\draw[->] (cav.south east) -- (cool.west);

% beam path
\node[box,below=1.9cm of cpl,minimum width=8.5cm,minimum height=0.9cm] (beamline) {Beamline Axis (beam aperture) \\ alignment \& vacuum};
\draw[->] (beamline.west) -- ++(-1.2,0) node[left,lab]{from linac};
\draw[->] (beamline.east) -- ++(1.2,0) node[right,lab]{to screen / spectrometer};

% notes
\node[lab,align=left,below=0.8cm of beamline]
  {Key design knobs: shunt impedance, coupling factor, breakdown limit, thermal stability,\\
   aperture vs.\ wakefields, mechanical alignment, LLRF phase noise, available RF power.};

\end{tikzpicture}%
}
\caption{RF feeding and integration of a TDS: the RF source (klystron or solid-state) delivers power through waveguides and phase control to a matched input coupler exciting the TM$_{110}$ deflecting mode. Higher-order mode (HOM) damping and active cooling maintain stability, while the beamline axis sets constraints on aperture, alignment, and vacuum.}
\label{fig:tds_rf_chain}
\end{figure}
\paragraph{Design Implications.} Mechanical and thermal stability strongly affect the mechanical support, cooling design, and alignment of the cavity. As the RF frequency increases, mechanical tolerances become tighter. High-frequency cavities are more compact and therefore more sensitive to vibrations, temperature variations, and alignment errors. For this reason, RF design, beam dynamics, and mechanical engineering must be considered together from the beginning. Only with this integrated approach can the expected temporal resolution be reliably achieved in routine operation.

These considerations are particularly important for free-electron lasers (FELs), plasma-based accelerators, and advanced light sources, where precise control of the longitudinal beam structure directly determines performance.

Although TDS technology is well established, significant challenges remain. Sub-femtosecond resolution requires high transverse voltage, excellent RF phase stability, and beam optics carefully matched to both the cavity and the spectrometer. At higher RF frequencies, mechanical tolerances become stricter, cavity apertures become smaller, and wakefield effects become more pronounced, which can distort the measurement.
Furthermore, optimizing temporal resolution and energy resolution at the same time introduces trade-offs in lattice design, dispersion settings, and screen placement.

 Fig.~\ref{fig:tds_rf_chain}illustrates the integration of the TDS system, including RF power delivery, mode excitation, beam transport, and thermal stabilization, into a tightly coupled architecture.

%%%%%%%%%%%%%%%%%%%%%%%%%%%
\section{Beamline Optics}

The achievable temporal and energy resolution of a TDS-based diagnostic system is determined not only by the RF deflecting voltage, but also by the beam transport optics between the cavity and the observation screen. 
Proper lattice design ensures that the transverse streaking signal is maximized while preserving beam quality and minimizing chromatic distortions.

\subsection{Optics Optimization for Temporal Resolution}

In the streaking plane, the transverse displacement at the screen is governed by the transport matrix element

\[
R_{12} = \sqrt{\beta_s \beta_0}\,\sin\Delta\psi,
\]

where $\beta_0$ and $\beta_s$ are the beta functions at the TDS and screen, respectively, and $\Delta\psi$ is the betatron phase advance between them. 
Since the temporal resolution scales inversely with $R_{12}$, maximizing this transport term is a primary design objective.

\paragraph{Phase Advance.}
The transport element $R_{12}$ reaches its maximum when the phase advance between the cavity and screen satisfies $\Delta\psi \approx 90^\circ$. 
Under this condition, the streak induced at the cavity is converted into maximum transverse displacement at the screen, providing optimal time-to-space mapping.

\paragraph{Beta Function at the Cavity.}
The intrinsic beam size inside the cavity scales with $\sqrt{\varepsilon \beta_0}$. 
Reducing $\beta_0$ in the streak plane decreases the transverse beam size within the structure, thereby increasing the ratio of streak-induced displacement to natural beam size. 
This improves the effective temporal resolution while also reducing sensitivity to cavity aperture limitations.

\paragraph{Control of Dispersion.}
Residual dispersion in the streak plane introduces an energy-dependent transverse offset, which can blur the time measurement in the presence of correlated or uncorrelated energy spread. 
Therefore, the horizontal dispersion $\eta_s$ at the screen must be minimized in the streaking plane to avoid mixing longitudinal phase space information.

\paragraph{Optical Stability.}
While maximizing $R_{12}$ enhances temporal resolution, excessively large beta functions at the screen can lead to increased beam size and sensitivity to chromatic aberrations. 
A balanced lattice design is therefore required to avoid excessive beam growth or nonlinear transport effects that degrade measurement accuracy.

\subsection{Optics Optimization for Energy Resolution}

When the TDS is combined with a downstream dipole spectrometer, the orthogonal transverse plane is used to resolve the beam energy distribution. 
In this configuration, the achievable energy resolution is limited by the transverse beam size at the screen and the dispersion function:

\[
\sigma_{\delta}^{\mathrm{res}} = \frac{\sigma_{y,0}}{D},
\]

where $D$ is the vertical dispersion at the screen and $\sigma_{y,0}$ is the intrinsic vertical beam size excluding dispersive contributions.

\paragraph{Large Dispersion at the Screen.}
Increasing the dispersion $D$ enhances the transverse separation per unit relative energy deviation, thereby improving energy resolution. 
However, large dispersion also increases sensitivity to higher-order chromatic effects and magnet nonlinearities, which must be carefully controlled.

\paragraph{Small Beta Function at the Screen.}
The intrinsic vertical beam size scales as $\sqrt{\varepsilon_y \beta_{y,s}}$. 
Reducing $\beta_{y,s}$ at the observation screen minimizes the betatron contribution to the measured beam size, lowering the resolution floor for energy measurements.

\paragraph{Plane Decoupling.}
To avoid mixing time and energy information, the lattice must maintain orthogonality between the streaking and dispersive planes. 
In practice, this requires suppressing dispersion in the streak plane ($\eta_x \approx 0$) and avoiding transverse coupling between horizontal and vertical optics.

\subsection{Energy-Measurement Specific Considerations}

Accurate reconstruction of the longitudinal phase space requires careful calibration and systematic corrections.

\paragraph{Dispersion Calibration.}
The dispersion $D$ should be determined through beam-based measurements, typically by scanning the linac phase to induce a controlled energy shift and recording the corresponding centroid displacement at the screen.

\paragraph{Imaging System Resolution.}
The finite point-spread function of the imaging system contributes an additional term $\sigma_R$ to the observed beam size. 
This contribution must be measured independently and subtracted in quadrature from the measured profile to obtain the true beam size.

\paragraph{Betatron Contribution.}
The intrinsic betatron size $\sqrt{\varepsilon_y \beta_{y,s}}$ must be characterized to separate geometric beam size from dispersive broadening. 
Independent measurements of emittance and beta function are therefore essential for reliable slice energy spread extraction.

\paragraph{Two-Scan Method.}
A commonly employed technique to disentangle intrinsic slice energy spread from TDS-induced correlations involves performing two systematic scans: varying the beam energy and varying the TDS voltage. 
This approach allows separation of geometric, dispersive, and RF-induced contributions to the observed beam profile.

Overall, the lattice design must ensure that temporal and energy resolutions are optimized simultaneously while preserving clear decoupling between the streaking and dispersive planes.

%%%%%%%%%%%%%%%%%%%%%%%%%%%%
\paragraph{Summary box.}
\begin{center}
\noindent\fbox{%
\parbox{0.94\linewidth}{%
In summary, TDS design proceeds iteratively. 
An initial RF frequency is selected based on the required temporal resolution. 
Beam optics are then optimized to maximize the effective shear parameter while preserving transverse emittance. 
Finally, stability tolerances (RF phase, amplitude, alignment) are evaluated to ensure feasibility under realistic operational conditions.

}}
\end{center}
%%%%%

%%%%%%%%%%%%%%%%%%%%%%%%%%%%%%
\section{Application to DALI}
\label{sec:dali}

This section applies the previously derived TDS performance relations to the DALI beam parameters at a kinetic energy of 50~MeV. 
The objective is to determine achievable temporal and energy resolution under realistic optics conditions and to evaluate the suitability of S-, C-, and X-band operation.

\begin{figure}
    \centering
    \includegraphics[width=0.8\linewidth]{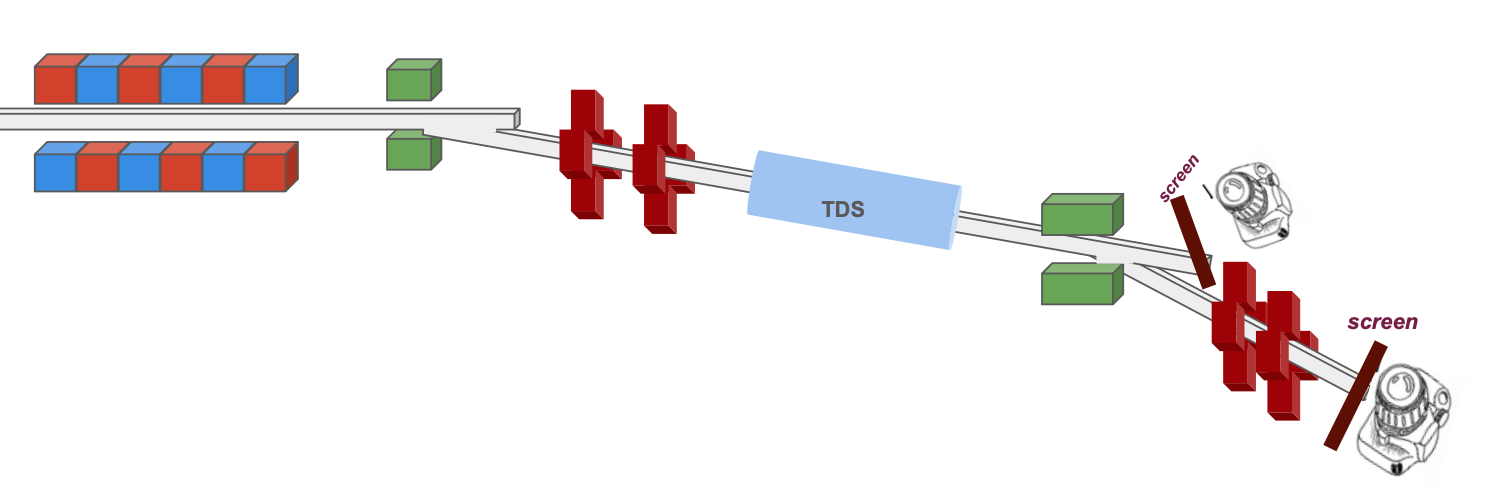}
    \caption{layout of the TDS after MId-FIR FEL }
    \label{fig:layout}
\end{figure}

The proposed installation location downstream of the Mid-FIR FEL is shown in Fig.~\ref{fig:layout}. 
The integration must satisfy aperture, optics, and space constraints while preserving beam quality for downstream transport.

\subsection{DALI Beam Parameters}

The following working parameters are assumed:

\begin{align*}
E_{\mathrm{kin}} &= 50~\mathrm{MeV}, \\
\sigma_t &\in [100,\,500]~\mathrm{fs}, \\
\varepsilon_x &= 5.52\times10^{-8}~\mathrm{m\,rad}, \\
\varepsilon_y &= 4.69\times10^{-8}~\mathrm{m\,rad}, \\
\beta_x &= 16.61~\mathrm{m}, \\
\beta_y &= 12.33~\mathrm{m}, \\
\sigma_\delta^{\mathrm{proj}} &\approx 5\times10^{-3}.
\end{align*}

The total beam energy is $E_{\mathrm{tot}}=50.511$~MeV, corresponding to $\gamma\approx98.85$ and $p_0\approx50.5~\mathrm{MeV}/c$.

\paragraph{Unstreaked sizes at the current optics.}
Assuming negligible dispersion in the streak plane,
\[
\sigma_{x,0}=\sqrt{\varepsilon_x \beta_x}\approx 0.958~\mathrm{mm},\qquad
\sigma_{y,0}=\sqrt{\varepsilon_y \beta_y}\approx 0.760~\mathrm{mm}.
\]
These values determine the starting point for time and energy resolution, respectively.
%\subsection{Choice of TDS Band and Transverse Voltage}
%We illustrate with an \textbf{S-band} TDS (\(f_{\mathrm{RF}}=2.998~\mathrm{GHz}\), \(\omega=2\pi f\)) and three plausible transverse voltages:
%\[
%V_\perp \in \{10,\,20,\,30\}\ \mathrm{MV}.
%\]
%Comparable formulas apply for C-/X-band via the $\omega$ scaling.
Here we consistently use $p_0 = \gamma m_e c$ in SI units unless otherwise stated.

%%%%%%%%%%%%%%%%%%%%%%
\subsection{Optics Working Point}

The transport between the TDS and the observation screen is characterized by

\[
R_{12}=\sqrt{\beta_s\beta_0}\sin\Delta\psi.
\]

A practical working point for DALI is chosen as

\[
\beta_0=3.0~\mathrm{m}, \text{(at TDS)},\quad
\beta_s=16.6~\mathrm{m}, \text{(at screen)},\quad
\Delta\psi=90^\circ,
\]

yielding

\[
R_{12}\approx7.06~\mathrm{m}.
\]

This configuration maximizes streak transport while keeping the screen beam size manageable.
%////////////////////
\subsection{Temporal Resolution Performance}

The temporal resolution is given by

\[
\sigma_t^{\mathrm{res}}=\frac{\sigma_{x,0}}{|S|},
\qquad
S=\frac{R_{12}}{p_0}\frac{eV_\perp\omega}{c}.
\]

For an S-band system ($f_{\mathrm{RF}}=2.998$~GHz), the expected resolutions are

\[
\sigma_t^{\mathrm{res}} \approx
\begin{cases}
36~\mathrm{fs}, & V_\perp=10~\mathrm{MV},\\
18~\mathrm{fs}, & V_\perp=20~\mathrm{MV},\\
12~\mathrm{fs}, & V_\perp=30~\mathrm{MV}.
\end{cases}
\]

These values are significantly smaller than the shortest anticipated bunch length (100~fs), demonstrating that S-band operation already provides sufficient temporal resolving power for DALI.
%///-------------------------------
\subsection{Band Comparison and Screen Constraints}
%\subsection{Band comparison at $V_\perp=10$\,MV and $\Delta\psi=90^\circ$}

We use
\[
\sigma_t^{\mathrm{res}}
= \frac{c\,p_0}{e\,V_\perp\,\omega}\,
\frac{\sqrt{\varepsilon_x/\beta_0}}{|\sin\Delta\psi|}
,\qquad
S_t \equiv \frac{\partial x}{\partial t}
= R_{12}\,\frac{e\,V_\perp\,\omega}{c\,p_0},
\]
with $R_{12}=\sqrt{\beta_s\beta_0}\sin\Delta\psi$ and $\sigma_x^2=\sigma_{x,0}^2+(S_t\sigma_t)^2$.  
DALI inputs: $E_\text{kin}=50~\text{MeV}$ ($\gamma\!\approx\!98.85$), $p_0\!\approx\!\gamma m_ec\beta \simeq 2.70{\times}10^{-20}\,\text{kg\,m/s}$,  
$\varepsilon_x=5.5246{\times}10^{-8}\,\text{m\,rad}$, $\beta_0=3.0$\,m, $\beta_{s,x}=16.6117$\,m $\Rightarrow R_{12}\approx 7.06$\,m, and $V_\perp=10$\,MV. 

\begin{table}[h]
\centering
\caption{Temporal resolution and streak amplitudes at $V_\perp=10$\,MV and $\Delta\psi=90^\circ$ (DALI optics).}
\label{tab:band_compare_10MV}
\begin{tabular}{l c c c c}
%\toprule
Band & $f_{\rm RF}$ (GHz) & $\sigma_t^{\mathrm{res}}$ (fs) & $x_{\rm rms}$ at 100\,fs (mm) & $x_{\rm rms}$ at 500\,fs (mm) \\
%\midrule
S-band & 2.998  & 36.4 & 2.63 & 13.16 \\
C-band & 5.712  & 19.1 & 5.02 & 25.08 \\
X-band & 11.424 & 9.55 & 10.03 & 50.16 \\
%\bottomrule
\end{tabular}
\end{table}

%\noindent
%\textit{Notes and scaling:}
%(i) $\sigma_t^{\mathrm{res}}\propto 1/\omega$ and $\propto \sqrt{1/\beta_0}$; reducing $\beta_0$ or increasing frequency boosts time resolution.
%(ii) $x_{\rm rms}=S_t\,\sigma_t$; the large streak at X-band may require reduced $R_{12}$ or $V_\perp$, or a larger field-of-view at the screen.
%(iii) If $\beta_0$ differs from 3\,m, rescale $\sigma_t^{\mathrm{res}}$ by $\sqrt{3\,\text{m}/\beta_0}$.
\begin{figure}
    \centering
    \includegraphics[width=0.8\linewidth]{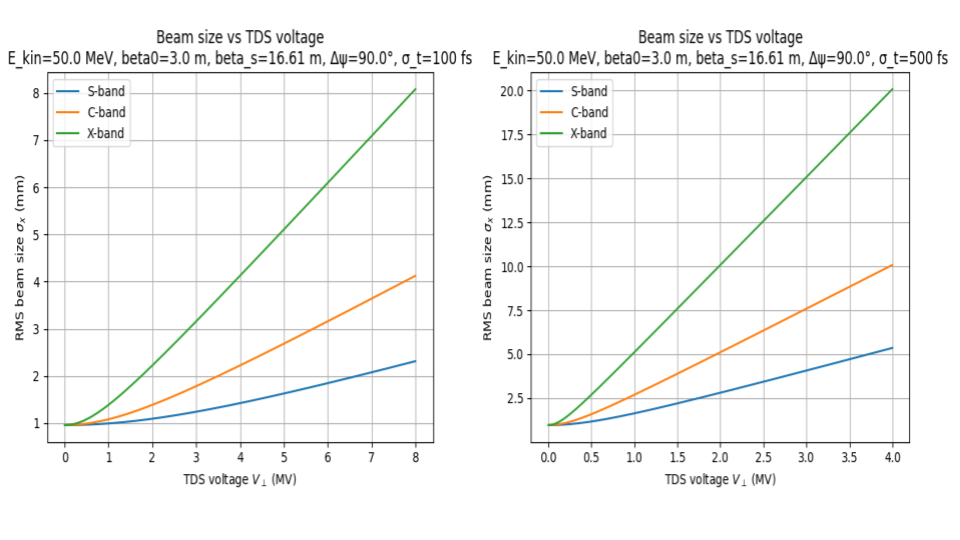}
    \caption{RMS beam size on the screen versus transverse voltage for (left) $\sigma_t=100$~fs and (Right) for $\sigma_t=500$~fs.}
    \label{fig:beam100fs}
\end{figure}

Figures~\ref{fig:beam100fs} left and right show the RMS beam size at the screen as a function of transverse voltage for S-, C-, and X-band operation, assuming the DALI optics defined above.
We already knew three important scaling behaviors from our formalist:(i) The streak amplitude grows linearly with transverse voltage; (ii) For identical voltage, higher RF frequency produces proportionally larger streak; (iii) The total beam size scales linearly with bunch length.

At X-band, the streak amplitude for long bunches (500~fs) can reach several tens of millimeters even at moderate voltage. 
Such large beam sizes may exceed practical screen field-of-view or introduce nonlinear imaging effects. 
While X-band offers superior intrinsic temporal resolution, it may be operationally excessive for the DALI bunch length regime.
\subsection{Energy Resolution}

In the vertical (dispersive) plane, the smallest resolvable slice energy spread is

\[
\sigma_\delta^{\mathrm{res}}=\frac{\sigma_{y,0}}{D},
\qquad
\sigma_{y,0}^2=\sigma_R^2+\varepsilon_y\beta_{y,s}.
\]

For the current optics, the betatron term dominates the vertical beam size. 
Reducing $\beta_{y,s}$ and increasing dispersion $D$ are therefore the most effective levers for improving energy resolution.

Because \(\varepsilon_y\beta_{y,s}\) dominates at the current optics, reducing \(\beta_{y,s}\) at the screen and increasing \(D\) are the most effective levers. The table below shows realistic scenarios (assuming \(\sigma_R{=}30~\mu\mathrm{m}\)):

\begin{center}
\begin{tabular}{c c c c}
%\toprule
$D$ (m) & $\beta_{y,s}$ (m) & $\sigma_{y,0}$ (µm) & $\sigma_{\delta}^{\mathrm{res}}$ \\
%\midrule
0.30 & 12.33 & 761 & $2.54\times10^{-3}$ \\
0.30 & 3.00  & 376 & $1.25\times10^{-3}$ \\
0.30 & 1.00  & 219 & $7.29\times10^{-4}$ \\
0.50 & 1.00  & 219 & $4.37\times10^{-4}$ \\
0.50 & 0.50  & 156 & $3.12\times10^{-4}$ \\
%\bottomrule
\end{tabular}
\end{center}

\noindent
\textit{Takeaway:} moving the screen optics toward \(\beta_{y,s}\!\sim\!1~\mathrm{m}\) and providing \(D\!\sim\!0.5~\mathrm{m}\) yields an energy-resolution floor of $(3–5\times 10^{-4})$ (i.e. $(\sim\!0.03–0.05\%))$, which is well below the projected \(\sigma_\delta^{\mathrm{proj}}\approx 5\times10^{-3}\) and thus suitable for resolving slice energy and slice spread.

%%%%%%%%%%%%%%%%%%%%%%%%%%%%%

% ------------------------------------------------------------------
% ------------------------------------------------------------------
% ------------------------------------------------------------------
% ------------------------------------------------------------------

% ------------------------------------------------------------------
\subsection{Overall Assessment for DALI}

From the combined temporal and energy analysis:

\begin{itemize}
    \item S-band operation at $V_\perp=20$–30~MV provides temporal resolution of 12–18~fs.
    \item The achievable energy-resolution floor is $\sim4\times10^{-4}$ with optimized optics.
    \item Higher-frequency operation improves intrinsic time resolution but leads to very large screen beam sizes.
\end{itemize}

Given the expected bunch length range of 100–500~fs and the moderate beam energy of 50~MeV, 
S-band operation represents the most balanced solution in terms of resolution, wakefield sensitivity, alignment tolerance, RF stability requirements, and screen dynamic range.

An S-band TDS installed downstream of the Mid-FIR FEL at DALI with

\[
f_{\mathrm{RF}}=2.998~\mathrm{GHz}, 
\quad
V_\perp=20–30~\mathrm{MV}, 
\quad
\beta_0=3~\mathrm{m}, 
\quad
\Delta\psi=90^\circ
\]
is therefore capable of delivering high-fidelity longitudinal phase space diagnostics 
suitable for both commissioning and routine DALI operation.

%%%%%%%%%%%%%%

\bibliographystyle{unsrt}
\bibliography{TDS}

@techreport{Emma2001,
  author       = {Emma, P. J.},
  title        = {A Transverse RF Deflecting Structure for Bunch Length and Phase Space Diagnostics},
  institution  = {Stanford Linear Accelerator Center, Menlo Park, CA (US)},
  doi          = {10.2172/784939},
  url          = {https://www.osti.gov/biblio/784939},
  place        = {United States},
  year         = {2001},
  month        = {06}}

@article{Krejcik:556141,
      author        = "Krejcik, P. and Akre, R. and Bentson, L. and Emma, P.",
      title         = "{A Transverse RF Deflecting Structure for Bunch Length and
                       Phase Space Diagnostics}",
      year          = "2001",
      url           = "https://cds.cern.ch/record/556141",
}

@article{Polarix2020,
  title = {Novel X-band transverse deflection structure with variable polarization},
  author = {Craievich, P. and Bopp, M. and others},
  journal = {Phys. Rev. Accel. Beams},
  volume = {23},
  issue = {11},
  pages = {112001},
  numpages = {22},
  year = {2020},
  month = {Nov},
  publisher = {American Physical Society},
  doi = {10.1103/PhysRevAccelBeams.23.112001}}

@article{Polarix2024,
  title = {Beam-based commissioning of a novel $X$-band transverse deflection structure with variable polarization},
  author = {Gonz\'alez Caminal, P. and Christie, F. and others},
  journal = {Phys. Rev. Accel. Beams},
  volume = {27},
  issue = {3},
  pages = {032801},
  numpages = {19},
  year = {2024},
  month = {Mar},
  publisher = {American Physical Society},
  doi = {10.1103/PhysRevAccelBeams.27.032801}}

@article{Behrens2014,
  author       = {Behrens, C. and Decker, F. -J. and and others},
  title        = {Few-femtosecond time-resolved measurements of X-ray free-electron lasers},
  doi          = {10.1038/ncomms4762},
  url          = {https://www.osti.gov/biblio/1131467},
  journal      = {Nature Communications},
  issn         = {ISSN 2041-1723},
  volume       = {5},
  place        = {United States},
  publisher    = {Nature Publishing Group},
  year         = {2014},
  month        = {07}}

@inproceedings{LOLA2010,
  author       = {Schreiber, S. and Faatz, B. and Feldhaus, J. and Honkavaara, K. and Treusch, R. and Vogt, M. and Rossbach, J.},
  title        = {FLASH Upgrade and First Results},
  booktitle    = {Proceedings of FEL 2010},
  series       = {Free Electron Laser Conference},
  year         = {2010},
  address      = {Malm\"o, Sweden},
  paper        = {TUOBI2},
  organization = {DESY}
}

@article{EGO2015381,
title = {Transverse C-band deflecting structure for longitudinal electron-bunch-diagnosis in XFEL “SACLA”},
journal = {Nuclear Instruments and Methods in Physics Research Section A: Accelerators, Spectrometers, Detectors and Associated Equipment},
volume = {795},
pages = {381-388},
year = {2015},
issn = {0168-9002},
doi = {https://doi.org/10.1016/j.nima.2015.06.018},
url = {https://www.sciencedirect.com/science/article/pii/S0168900215007639},
author = {H. Ego and H. Maesaka and T. Sakurai and Y. Otake and T. Hashirano and S. Miura},
}

@article{PSI_CBANC,
    author = {Prat, E. and Malyzhenkov, A. and Craievich, P.},
    title = {Sub-femtosecond time-resolved measurements of electron bunches with a C-band radio-frequency deflector in x-ray free-electron lasers},
    journal = {Review of Scientific Instruments},
    volume = {94},
    number = {4},
    pages = {043103},
    year = {2023},
    month = {04},
    issn = {0034-6748},
    doi = {10.1063/5.0144876},
    url = {https://doi.org/10.1063/5.0144876},
    eprint = {https://pubs.aip.org/aip/rsi/article-pdf/doi/10.1063/5.0144876/17455701/043103_1_5.0144876.pdf},
}

@techreport{Loew1965RFDeflecting,
  author       = {Loew, G. A. and Altenmueller, O. H.},
  title        = {Design and Applications of R.F. Deflecting Structures at SLAC},
  institution  = {Stanford Linear Accelerator Center},
  address      = {Stanford, California, USA},
  year         = {1965},
  note         = {Presented by G. A. Loew}
}

@article{Eduard2020,
  title = {High-resolution dispersion-based measurement of the electron beam energy spread},
  author = {Prat, E. and Dijkstal, P. and Ferrari, Eugenio. and Malyzhenkov, A. and Reiche, S.},
  journal = {Phys. Rev. Accel. Beams},
  volume = {23},
  issue = {9},
  pages = {090701},
  numpages = {9},
  year = {2020},
  month = {Sep},
  publisher = {American Physical Society},
  doi = {10.1103/PhysRevAccelBeams.23.090701},
  url = {https://link.aps.org/doi/10.1103/PhysRevAccelBeams.23.090701}
}

@article{FLASH2009,
  title = {Time-resolved electron beam phase space tomography at a soft x-ray free-electron laser},
  author = {R\"ohrs, Michael and Gerth, Christopher and Schlarb, Holger and Schmidt, Bernhard and Schm\"user, Peter},
  journal = {Phys. Rev. ST Accel. Beams},
  volume = {12},
  issue = {5},
  pages = {050704},
  numpages = {13},
  year = {2009},
  month = {May},
  publisher = {American Physical Society},
  doi = {10.1103/PhysRevSTAB.12.050704},
  url = {https://link.aps.org/doi/10.1103/PhysRevSTAB.12.050704}
}

@book{Wangler2008,
  author    = {Thomas P. Wangler},
  title     = {RF Linear Accelerators},
  publisher = {Wiley-VCH},
  year      = {2008},
  edition   = {2},
  isbn      = {978-3527406807}
}

@book{ChaoMess2013,
  editor    = {Alexander Wu Chao and Maury Tigner and Frank Zimmermann},
  title     = {Handbook of Accelerator Physics and Engineering},
  publisher = {World Scientific},
  year      = {2013},
  edition   = {2},
  isbn      = {978-9814415859}
}

@book{Lee2018,
  author    = {S. Y. Lee},
  title     = {Accelerator Physics},
  publisher = {World Scientific},
  year      = {2018},
  edition   = {4},
  isbn      = {978-9813278738}
}

@incollection{Palumbo1994,
  author    = {L. Palumbo and V. G. Vaccaro and M. Zandbergen},
  title     = {Wakefields and Impedance},
  booktitle = {Handbook of Accelerator Physics and Engineering},
  editor    = {A. W. Chao and M. Tigner},
  publisher = {World Scientific},
  year      = {1999}
}

@article{Akre2002,
  author  = {R. Akre and others},
  title   = {Bunch length measurements using a transverse RF deflecting structure in the SLAC FFTB},
  journal = {Proceedings of EPAC 2002},
  year    = {2002}
}

@techreport{CERNRFSchool,
  author      = {CERN Accelerator School},
  title       = {RF Engineering for Particle Accelerators},
  institution = {CERN},
  year        = {2012}
}

@article{Schlarb2013,
  author  = {H. Schlarb and others},
  title   = {The European XFEL timing system},
  journal = {Proceedings of FEL 2013},
  year    = {2013}
}

@article{FLASHTDS2009,
  title = {Time-resolved electron beam phase space tomography at a soft x-ray free-electron laser},
  author = {R\"ohrs, Michael and Gerth, Christopher and Schlarb, Holger and Schmidt, Bernhard and Schm\"user, Peter},
  journal = {Phys. Rev. ST Accel. Beams},
  volume = {12},
  issue = {5},
  pages = {050704},
  numpages = {13},
  year = {2009},
  month = {May},
  publisher = {American Physical Society},
  doi = {10.1103/PhysRevSTAB.12.050704},
  url = {https://link.aps.org/doi/10.1103/PhysRevSTAB.12.050704}
}

@article{Floettmann,
  title = {Beam dynamics in transverse deflecting rf structures},
  author = {Floettmann, Klaus and Paramonov, Valentin V.},
  journal = {Phys. Rev. ST Accel. Beams},
  volume = {17},
  issue = {2},
  pages = {024001},
  numpages = {11},
  year = {2014},
  month = {Feb},
  publisher = {American Physical Society},
  doi = {10.1103/PhysRevSTAB.17.024001},
  url = {https://link.aps.org/doi/10.1103/PhysRevSTAB.17.024001}
}

\end{document}